\documentclass[twocolumn]{aastex631}
\usepackage{float}
\makeatletter
\let\newfloat\newfloat@ltx
\makeatother
\usepackage{algorithm}
\usepackage{algpseudocode}
\usepackage{comment}
\usepackage{todonotes}
\usepackage{amsmath}
\usepackage{hyperref}
\usepackage{CJKutf8}

\newcommand{\vbbl}{\texttt{VBBinaryLensing}}
\newcommand{\jaxcode}{\texttt{microlux}} %

\begin{document}

\title{A differentiable binary microlensing model using adaptive contour integration method}

\author[0009-0000-7721-6342]{Haibin Ren\begin{CJK*}{UTF8}{gbsn} (任海滨)\end{CJK*}}

\affiliation{Department of Astronomy, Tsinghua University, Beijing 100084, China}

\author[0000-0003-4027-4711]{Wei Zhu\begin{CJK*}{UTF8}{gbsn} (祝伟)\end{CJK*}}
\affiliation{Department of Astronomy, Tsinghua University, Beijing 100084, China}

\correspondingauthor{Wei Zhu}
\email{weizhu@tsinghua.edu.cn}

\begin{abstract}

We present \jaxcode, which is a \texttt{Jax}-based code that can compute the binary microlensing light curve and its derivatives both efficiently and accurately.
The key feature of \jaxcode\ is the implementation of a modified version of the adaptive sampling algorithm that was originally proposed by V.\ Bozza to account for the finite-source effect most efficiently.
The efficiency and accuracy of \jaxcode\ have been verified across the relevant parameter space for binary microlensing. 
As a differentiable code, \jaxcode\ makes it possible to apply gradient-based algorithms to the search and posterior estimation of the microlensing modeling.
As an example, we use \jaxcode\ to model a real microlensing event and infer the model posterior via both Fisher information matrix and Hamiltonian Monte Carlo, neither of which would have been possible without the access to accurate model gradients.

\end{abstract}


\section{Introduction} \label{sec:intro}

Gravitational microlensing has been successful in detecting exoplanets and stellar binaries \citep{mao1991, gould1992}. According to the NASA Exoplanet Archive \citep{Akeson:2013},
\footnote{Based on a query made in December 2024.}
microlensing has discovered more than 200 planets, many of which have relatively wide orbits and/or low planet-to-star mass ratios. This makes microlensing one promising method for finding Earth-like exoplanets \citep[e.g.,][]{Bennett:1996, penny2019} and complementing the other detection techniques in studying exoplanet demographics \citep[e.g.,][]{gaudi2012,zhu2021}. 

The interpretation of planetary and binary-star microlensing (hereafter binary microlensing both together) events requires algorithms that can efficiently and accurately compute the microlensing light curve. When the source star is sufficiently away from the caustic curve, which is a set of points with infinite magnification, the source is effectively point-like and the microlensing magnification can be computed quite easily. However, when the source becomes close enough to the caustic, the finite size of the source cannot be ignored, and the computation of the magnification involves a two-dimensional integration over the projected surface of the source.
Several numerical algorithms have been made available to tackle this problem, including inverse ray shooting \citep{kayser1986}, contour integral \citep{schramm1987,gould1997,dominik1998,bozza2010}, and hybrid algorithms that combine both \citep{dong2006, dominik2007}. In particular, the contour integral method simplifies and accelerates the calculations by transforming the areal integral into one dimension via Green's theorem. First introduced by \citet{schramm1987}, this method has been extensively used in binary microlensing \citep[e.g.,][and thereafter]{gould1997, dominik1998}.

One of the most widely used algorithms in computing the binary microlensing light curve is \vbbl\ \citep{bozza2010,bozza2018}.
\footnote{This package is now part of the new package \texttt{VBMicrolensing} \citep{bozza2024a}. The key functions related to binary microlensing remain unchanged.}
By controlling serval specifically designed error estimators, \vbbl\ can perform adaptive sampling of the contour integration method and achieve smooth transition from the point-source approximation. This package has found extensive use in the modelings of many binary and planetary events \citep[e.g.,][]{Zang:2021} as well as several open-source packages \citep[e.g.,][]{poleski2019, bachelet2017}. 

With the diversified goals of microlensing studies as well as the upcoming new era of microlensing surveys, there is the need for new methods and frameworks in microlensing modeling. The existing surveys, most notably OGLE \citep[Optical Gravitational Lensing Experiment,][]{udalski2003}, MOA \citep[Microlensing Observations in Astrophysics,][]{bond2001}, and KMTNet \citep[Korea Microlensing Telescope Network,][]{kmtnet2016}, have altogether detected $>10^5$ microlensing events, and the next generation of microlensing surveys are promised to detect even more events with higher quality \citep{penny2019, yan2022}. Within these large datasets, events that are intrinsically rare occur, including those with dark lenses \citep{sahu2022, lam2022, mroz2022}, lenses of multiple ($\geq3$) components \citep[e.g.,][]{gaudi2008, gould2014}, etc. While some clever search strategy can be adopted to identify some of those intrinsically rare but astrophysically interesting events, an improved modeling method and framework will almost certainly uncover more unexpected microlensing phenomena and allow us to study them in a more systematic way.

New tools on scientific programming have been made available especially driven by the advancement of the machine learning field.
In particular, several modern packages have been developed that can automatically build the computational graph and apply the chain rule to obtain the model derivatives up to machine precision (i.e., automatic differentiation; e.g., \texttt{Julia}, \citealt{Julia-2017}, \texttt{Jax}, \citealt{jax2018github}). These modern packages use many new techniques that have improved the applicability and performance of the automatic differentiation method \citep{autodiff2018}.
There have been successful applications of this method in probabilistic programming \citep[e..g, \texttt{Stan},][]{carpenter2017}, machine learning \citep[e.g., \texttt{PyTorch},][]{Ansel_PyTorch_2_Faster_2024}, and many other fields.
The modeling procedure of microlensing data may also be improved with the access to accurate model derivatives.

In this work, we introduce \jaxcode, a \texttt{Jax}-based code for magnification and gradient calculations of binary microlensing. Compared to another similar attempt \cite{bartolic2023a}, our work focuses on the implementation of the adaptive sampling method of \citet{bozza2010} and can produce model gradients in an efficient and accurate manner. This paper is organized in the following way: Section~\ref{sec:overview} provides a brief overview of the adaptive sampling method and its implementation in \vbbl; Section~\ref{sec:contour_in_jax} explains in details how the method is modified and implemented in \texttt{Jax}; 
Section~\ref{sec:performance} provides tests that validate \jaxcode\ and evaluate its performance; Section~\ref{sec:light_curve_analysis} demonstrates the utility of \jaxcode\ by applying it to the analysis of a real, archival event; finally Section~\ref{sec:discussion} provides a brief summary of the present work and short discussions about potential future applications and improvements.

\section{Overview of \vbbl} \label{sec:overview}

For any point on the source plane,
its corresponding images can be obtained by solving the binary lens equation
\begin{equation} \label{eqn:lens-eq}
        \zeta = z - \frac{1/(1+q)}{\bar{z} - \bar{z}_1} - \frac{q/(1+q)}{\bar{z} - \bar{z}_2} .
\end{equation}
Here $\zeta = y_1+iy_2$ is the complex notation of the source position, $q$ is the mass ratio, $z=x_1+ix_2$ is the image position, $\bar{z}$ is its complex conjugate, and $\bar{z}_1$ and $\bar{z}_2$ are the complex conjugates of the binary lens positions, respectively \citep{witt1990}. In \vbbl, the secondary lens is fixed at origin (i.e., $z_2=0$), and the primary lens is on the real axis at $-s$.

The binary lens equation (Equation~\ref{eqn:lens-eq}) can be converted into a fifth-order complex polynomial and then solved numerically \citep{witt1990}. \vbbl\ applies the improved Laguerre method of \citet{skowron2012} and identifies all five solutions in an iterative way, which are later polished using the original polynomial to reduce the numerical error. Each solution has also an associated parity $p$, which is the sign of the determinant of the Jacobian \citep{witt1995}
\begin{equation}
    \det J \equiv 1-\left| \frac{\partial \zeta}{\partial \bar{z}} \right|^2 .
\end{equation}
The solutions are then substituted into the original lens equation to eliminate the spurious solutions. In doing so, \vbbl\ makes use of the known facts that (1) there are either three or five solutions of the binary lens equation, depending on whether the source is outside or inside the caustic, and (2) the sum of the parity of all true solutions should be $-1$ \citep{witt1995}. In addition, \vbbl\ also applies some empirical criteria related to the relative accuracy of the third and fourth worse solutions.

For a typical binary microlensing light curve, the point source approximation by summing up the inverse Jacobian determinant for each true image serves well for the majority of the time. To identify those source positions that require special treatment for the finite-source effect, \vbbl\ applies a number of tests to speed up the light curve calculation significantly \citep{bozza2018}. At any source position where the point source approximation fails, \vbbl\ selects a sample of points on the boundary of the source circle
\begin{equation}
    \zeta = \zeta_{c} + \rho e^{i\theta_i}, \theta \in [0,2\pi]
\end{equation}
and solves the lens equation for their corresponding image positions. Here $\zeta_c$ is the position of the source center, $\rho$ is the radius of the source normalized to the Einstein ring radius $\theta_{\rm E}$. In order to compute the total magnification, \vbbl\ computes the total area enclosed by the image points, using the numerical approximation of the Green's theorem
\begin{equation}
    A_{\rm image} = \sum_{I} p_I \sum_i  ( \Delta A_{I,i}^{\rm (t)} + \Delta A_{I,i}^{\rm (p)} ) .
\end{equation}
Here $p_I$ is the parity of the $I$-th image, $\Delta A_{I, i}^{\rm (t)}$ is the trapezoidal approximation of the line integral from the argument interval $(\theta_i, \theta_i+\Delta \theta)$,
\begin{equation}
    \Delta A_{I,i}^{\rm (t)} = \frac{1}{2} \boldsymbol{x}_I(\theta_i) \wedge \boldsymbol{x}_I (\theta_i + \Delta \theta) ,
\end{equation}
and $\Delta A_{I, i}^{\rm (p)}$ is the parabolic correction
\begin{equation} \label{eqn:parabolic}
    \Delta A_{I,i}^{\rm (p)} = \frac{1}{24} [(\boldsymbol{x}'_I \wedge \boldsymbol{x}''_I)|_{\theta_i} + (\boldsymbol{x}'_I \wedge \boldsymbol{x}''_I)|_{\theta_i+\Delta \theta}] \Delta \theta^3 .
\end{equation}
The symbol $\boldsymbol{x}_I(\theta_i)$ denotes the $I$-th image position vector of the point on the source boundary with angle $\theta_i$, and $\boldsymbol{x}'_I(\theta_i)$ is its derivative with respect to $\theta$.

In the core of \vbbl\ is its strategy to adaptively sample points on the boundary of the source circle according to the requirement magnification precision \citep{bozza2010}. 
There are a few notable features in this adaptive sampling process. First, \vbbl\ starts with two points and every iteration adds one new sampling to the source arc that contributes largest to the magnification error. There is in principle no limit on the maximum number of sampling points.
Second, once a new point is added, the corresponding image positions are connected to the closest points on the existing image polygons with the same parity. Third, the total error estimator, given first in \citet{bozza2010} and recently updated in \cite{bozza2024a}, concerns only the estimated error on the microlensing magnification.

To deal with the limb darkening effect, \vbbl\ divide the source into several concentric annuli with different constant average brightness according to the limb darkening profile. The total magnification is the weighted sum of the uniform brightness magnification on these annuli. Similar to the adaptive sampling, \vbbl\ also use several error estimators to optimize the division until the total error reach the desired accuracy.

\section{Realization of \jaxcode}\label{sec:contour_in_jax}

The implementation of the adaptive sampling algorithm in \texttt{Jax} is not a simple translation of \vbbl\ from \texttt{C++} into \texttt{Jax}. This is because \vbbl\ makes use of the flexible data structures such as the linked list and a number of dynamical algorithms in \texttt{C++}. As an array-oriented and vectorized programming language, \texttt{Jax} works the best with arrays with fixed shapes. Because of the differences between the two languages, a number of changes need to be made to realize the adaptive sampling method, as will be detailed later in this section. The advantage of using \texttt{Jax} is of course the automatic differentiation, and we show in Section~\ref{sec:Automatic differentiation} how it is achieved.

\subsection{Solving Lens Equation} \label{sec:roots_finding}

As mentioned in \citet{bozza2010}, most of the computing time in \vbbl\ is spent in solving the lens equation, so it is crucial to adapt a suitable root-finding algorithm. \vbbl\ uses the \citet{skowron2012} algorithm, which identifies solutions successively. %
In order to achieve high performance in \texttt{Jax}, the lens equation solver should be parallelized and vectorized. This calls for an algorithm that can identify all roots of the polynomial equation simultaneously.

We adopt the Aberth--Ehrlich method \citep{aberth1973,ehrlich1967}, which uses an implicit deflation strategy and can solve all the roots of polynomials simultaneously (see also \citealt{bartolic2023a}). 
The update formula is given by
\begin{equation}
    z_k^{\rm new} = z_k - \left( \frac{P^\prime (z_k)}{P(z_k)} - \sum_{j=1, j \neq k}^n (z_k - z_j)^{-1} \right)^{-1} .
\end{equation}
Here $P(z_k)$ and $P'(z_k)$ are the function value and the derivative of the polynomial with order $n$ evaluated at $z_k$, respectively, with $z_k$ the $k$-th root.%
The above update rule is similar to that of the standard Newton's method, except for the addition of the second term in the denominator. As long as the initial guesses are not off by a significant amount, the convergence of the Aberth--Ehrlich method is cubic 
\citep{aberth1973,ehrlich1967}
, similar to the Laguerre method that is commonly used for polynomial root findings.
We use the method proposed by \citet{fatheddin2022} to set the initial guess for the first point, and use the roots from the previous polynomial as the initial guess for the next, in order to achieve better performance.

To identify the real roots and get the correct parity, we make use of the two physical properties in the binary lens case and a relative tolerance criterion once the solutions are inserted back into the original lens equation in the same way as \vbbl\ (see Section~\ref{sec:overview}).

\subsection{Connecting Image Contours} \label{sec:contour}

Once the image positions are obtained, the task is then to connect these points into image contours. For regular cases, the basic rule is to connect two image points with the minimum distance and the same parity. When there are image creations (i.e., caustic entrance, with image number changing from three to five) or image destructions (i.e., caustic exit, with image number changing from five to three), there are extra points for certain source positions that need to be connected to each other.%

We use the Jonker--Volgenant algorithm \citep{Jonker1987ASA} to solve the linear sum assignment problem and connect all the image points.
The cost matrix that includes both the geometric distance and the parity information is given by
\begin{equation}
    C_{m,n,i} = |z_{i-1}[m] -z_{i}[n]| + K |p_{i-1}[m]-p_{i}[n]| ,
\end{equation}
where $C_{m,n,i}$ is the cost between the $m$-th image of the source position with $\theta_{i-1}$, $z_{i-1}[m]$, and the $n$-th image of the next source position with $\theta_i$, $z_{i}[n]$, and $p_{i-1}[m]$ and $p_{i}[n]$ are the corresponding parity values, respectively.
The parameter $K$ is a tunable constant to ensure the image match with same parity, and we find $K=5$ is a reasonable number. The cost matrix will be a square matrix of shape $(5, 5)$ or a rectangular matrix of shape either $(5, 3)$ (for image destruction) or $(3, 5)$ (for image creation). The latter cases are also called unbalanced assignments.

We implement the modified Jonker--Volgenant algorithm, which can handle the unbalanced assignments and is currently used in \texttt{scipy.optimize.linear\_sum\_assignment} \citep{scipy_lsa}, in \texttt{Jax}. 
This algorithm is guaranteed to find the global best match and thus robust, making it suitable for the problem of imaging connecting in our code.
Although the theoretical time complexity of the Jonker--Volgenant algorithm is $O(n^3)$, the actual executing time is short because the matching between the majority of the points is already satisfied when we use the roots from last $\theta$ sampling as the initial guess (see Section~\ref{sec:roots_finding}). 
Given these properties, we suggest that people consider this method in connecting the image contours in the triple-lens (and potentially higher multiples) case.

\subsection{Adaptive Sampling} \label{sec:adaptive sampling in jax}

Adding the sampling points one-by-one is very inefficient to fulfill optimal performance in \texttt{Jax} because of the frequent data movement. %
Given that the optimal sampling requires eight points at least \citep{bozza2018}, we choose to start with more points (30 by default, see Section~\ref{sec:performance}) and add multiple points in multiple intervals in a single iteration. Within each $\theta$ interval, the number of points to be added within the interval $(\theta_i, \theta_{i+1})$ is determined by
\begin{equation}\label{eqn:add_samplings}
    N_{\rm new,i} = {\rm min}(\left[ \left( \frac{E_i \sqrt{N}}{k \epsilon} \right) ^{\frac{1}{5}} \right], N_{\rm{max}} ) . 
\end{equation}
Here $E_i$ is the value of the error estimator in the certain $\theta$ interval ($\theta_i$ to $\theta_{i+1}$), $N$ is the current total number of points in this iteration, $\epsilon$ is the accuracy goal for the magnification calculation, $k$ is a parameter that is empirically chosen to transform the accuracy goal of total magnification into the accuracy in the given $\theta$ interval and is set to $k=2$ by default,
and $N_{\rm max}=4$ is also an empirical upper limit to avoid over-sampling when the error estimators explode.
The above formula is derived from the relative difference between the tolerance ($\epsilon$) and the current error ($E_i$), and the power law index chosen here is the order of the error estimator. %
With the above sampling strategy, we can then use vectorization to improve the performance of the code.

Similar to \vbbl\ \citep{bozza2018}, we also include the quadrupole test \citep{pejcha2009,gould2008,cassan2017}, ghost image test, and planetary test to identify the regions where the special treatment for the finite-source effect is needed.
\footnote{Note that there is inconsistency between some of the equations related to these tests in \citet{bozza2018} and their realizations in \vbbl (V.\ Bozza, private communication). These have been corrected in \jaxcode.} 

To account for the limb-darkening effect, we also use the weighted sum of magnifications from a number of concentric annuli, each with uniform surface brightness and contributing equally to the total source area. In its current implementation, \jaxcode\ follows the general convention and adopts the linear limb-darkening law \citep{milne1921}. The number of concentric annuli is set at 10 by default, which is sufficient for the general purpose \citep[e.g.,][]{dong2006}, although this number can be easily modified by the user.

\subsection{Using Arrays with Fixed Lengths}
\label{sec:bottleneck}

\texttt{Jax} provides users a \texttt{numpy}-like interface with a \texttt{C}-like execution performance via the Just-In-Time (JIT) compilation \citep{jax2018github}. 
However, to make use of JIT compilation, users are required to fix the shapes of arrays in the code, which is not naturally achieved in dynamical problems such as the adaptive sampling of microlensing.

Here we use a dynamic array-like structure and predefine multiple arrays of fixed and increasing lengths. 
Whenever contour integration is invoked, we start with the array of the shortest length, which is set to $30$ by default but adjustable by the user. This is sufficient to cover most cases that need the contour integral method (see Section~\ref{sec:magnification_test}). At each iteration of the adaptive sampling, we examine the total number of points against the length of the current array. If the current array is long enough, then the newly added points are assigned to the current array. Otherwise, all elements in the array 
are copied to the array of a longer length to avoid overflow. This data movement continues during the iterations of adaptive sampling, until the stopping criterion of the adaptive sampling is met or the total number of sampling points exceeds the maximum length (480 by default) of all arrays. In this latter situation, which is rarely encountered with our choices of the array lengths (see Section~\ref{sec:performance}), the adaptive sampling stops and a warning is issued. The magnification in the last iteration is chosen as the final result. 

Our above approach requires more memory and comes with longer JIT compilation time, but the price is affordable in order to enable fast execution and automatic differentiation.

\subsection{Automatic Differentiation}\label{sec:Automatic differentiation}
Once the code is written in \texttt{Jax}, it is generally straightforward to obtain automatic differentiation via functions like \texttt{jax.jacfwd} or \texttt{jax.jacrev}. However, it is not guaranteed that these functions will always produce the correct results. There are a few features in \jaxcode\ that require special treatments in order to have the automatic differentiation work properly.

First, for iterative algorithms like root-finding, the numerical instability may lead to \texttt{Nan} values in the returned derivatives. To avoid such a situation, we use the implicit function theorem to get the derivatives of the roots with respect to the polynomial coefficients.
Specifically, for a polynomial of order $n$
\begin{equation}
    f(a_k,z) = \sum_{k=0}^n a_k z^k ,
\end{equation}
where $\{a_k\}$ are the coefficients and $z$ the variable of the polynomial, the derivatives of one given root $z=z_0$ with respect to the polynomial coefficients are given as
\begin{equation}
     \frac{\partial z_0}{\partial a_k}  =\left. \left[ -\frac{\partial f}{ \partial a_k} / \frac{\partial f}{\partial z} \right] \right| _{z=z_0}  .
\end{equation}
This conversion can be achieved via \texttt{Jax.lax.custom\_root}. With this modification, the computational graph in automatic differentiation is also simplified. %

Second, for iterative algorithms like adaptive sampling, there is no need to trace the gradients inside all loops, and analytical results are available to accelerate the calculations and reduce the memory cost. We therefore stop the gradient tracing at the function that performs the adaptive sampling and use \texttt{Jax.custom\_jvp} to set up a custom JVP (Jacobian-vector product) rule. Following \citet{bozza2010}, the derivative of a source position $\zeta$ with respect to any model parameter $\Theta$ can be derived from Equation~(\ref{eqn:lens-eq})
\footnote{The following derivations assume the primary lens at $s$.}
\begin{equation}
    \frac{\partial \zeta}{\partial \Theta} = \frac{\partial z}{\partial \Theta} + \frac{\partial \zeta}{\partial \bar{z}} \frac{\partial \bar{z}}{\partial \Theta} + \frac{\partial q}{\partial \Theta} \frac{\partial \zeta}{\partial q} +\frac{\partial s}{\partial \Theta} \frac{\partial \zeta}{\partial s} \label{eqn:source_der} ,
\end{equation}
where
\begin{equation}
    \frac{\partial \zeta}{\partial \bar{z}} = \frac{1}{1+q}\left[\frac{1}{(\bar{z}-s)^2} + \frac{q}{\bar{z}^2}\right] ,
\end{equation}
\begin{equation}
    \frac{\partial \zeta}{\partial q} = \frac{1}{(1+q)^2}\left(\frac{1}{\bar{z}-s}-\frac{1}{\bar{z}} \right) ,
\end{equation}
and
\begin{equation}
    \frac{\partial \zeta}{\partial s} = -\frac{1}{1+q}\frac{1}{(\bar{z}-s)^2} .
\end{equation}
Together with its complex conjugate, Equation~(\ref{eqn:source_der}) yields
\begin{multline}
    \frac{\partial z}{\partial \Theta} = \left[ \frac{\partial \zeta}{\partial \Theta} - \frac{\partial \zeta}{\partial \bar{z}} \frac{\partial \bar{\zeta}}{\partial \Theta} - \frac{\partial q}{\partial \Theta}(\frac{\partial \zeta}{\partial q} - \frac{\partial \bar{\zeta}}{\partial \bar{q}} \frac{\partial \zeta}{\partial \bar{z}} ) \right. \\
    \left. - \frac{\partial s}{\partial \Theta}(\frac{\partial \zeta}{\partial s} - \frac{\partial \bar{\zeta}}{\partial \bar{s}} \frac{\partial \zeta}{\partial \bar{z}})
    \right] J^{-1}.
\end{multline}
With this custom JVP rule we can also achieve reverse mode automatic differentiation in the \texttt{while} loops, which otherwise is not achievable in \texttt{Jax} \footnote{\href{https://jax.readthedocs.io/en/latest/_autosummary/jax.lax.while_loop.html}{https://jax.readthedocs.io/en/latest/\_autosummary/\newline jax.lax.while\_loop.html}}.

Another issue is related to the adaptive step size in the numerical integration \citep{Eberhard1999}. 
Because the step size is adjusted based on the model parameters, it is also an implicit function of the model parameters. 
Directly applying the chain rule to the numerical integration with the parabolic correction, one would have
\begin{equation} \label{eqn:step-size}
    \frac{ \partial (\Delta A_{I,i}^{\rm (p)})}{\partial \Theta} =  \frac{ \partial f(\boldsymbol{x})}{\partial \Theta} \Delta \theta^3+ f(\boldsymbol{x}) \frac{ \partial (\Delta \theta^3)}{\partial \Theta} ,
\end{equation}
where $f(\boldsymbol{x}) \equiv  [(\boldsymbol{x}'_I \wedge \boldsymbol{x}''_I)|_{\theta_i} + (\boldsymbol{x}'_I \wedge \boldsymbol{x}''_I)|_{\theta_i+\Delta \theta}]/24 $ (Equation~\ref{eqn:parabolic}). While the second term on the right hand side of Equation~(\ref{eqn:step-size}) is introduced by the code, only the first term is needed for the correct derivatives.
To resolve the issue, we use \texttt{Jax.lax.stop\_gradient} to stop the gradient trace of the step size in the automatic differentiation.

Finally, the convergence of the function value in the adaptive sampling does not necessarily lead to the convergence of the derivatives.
This is especially true in binary lensing because of the existence of singularity at caustic. The experiment that will be shown in Section~\ref{sec:gradient} indicates that the derivatives of the numerical integration to the binary microlensing parameters become so large that the original error estimators do not promise the convergence of the model derivatives. To solve this problem, we introduce a new error estimator
\begin{equation} \label{eqn:new-error}
    E_{I,i}^{\rm d} =\frac{1}{240}\left| ( \boldsymbol{x}'_{I} \wedge \boldsymbol{x}'''_I)|_{\theta_i} - (\boldsymbol{x}'_{I} \wedge \boldsymbol{x}'''_I)|_{\theta_{i+1}} \right| \Delta \theta^3 .
\end{equation}
Here $\boldsymbol{x}'_{I} \wedge \boldsymbol{x}'''_I$ can be obtained from the analytical formula %
\begin{equation}
    \boldsymbol{x}' \wedge \boldsymbol{x}''' = \frac{\partial \rm{Im} \left[(z')^2 \zeta' \frac{\partial^2 \bar{\zeta}}{\partial z^2} J^{-1} \right]}{\partial \theta
    } .
\end{equation}
Similar to the first error estimator of \vbbl\ \citep{bozza2010}, this new error estimator will add more sampling points where the derivatives of the numerical integration (especially the parabolic correction item) are large.
Because the original error estimators are enough to ensure dense sampling around critical points and the calculations of these third derivatives are heavy, the new error estimator is only added in the ordinary images at caustic crossing.

\begin{figure*}[htb!]
    \includegraphics[width=\linewidth]{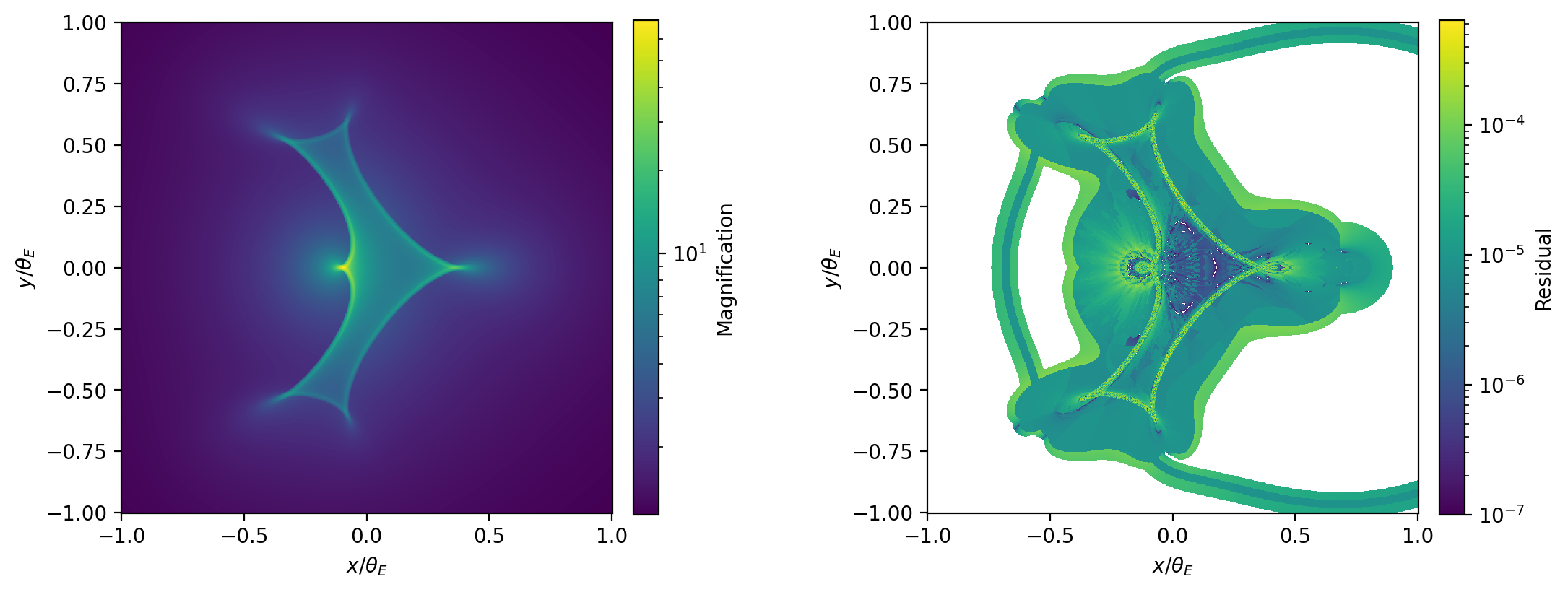}
    \caption{The left panel shows the magnification map produced by \jaxcode\ for a binary lens configuration with $s = 0.9$, $q = 0.2$, and $\rho = 10^{-3}$, and the right panel is the fractional deviation in magnification between \jaxcode\ and \vbbl. In both calculations, we have set both the absolute tolerance and the relative tolerance to $10^{-3}$. The two codes produce very similar magnification maps, and the small deviations arise from the regions where the quadrupole test fails, either in \vbbl\ or \jaxcode\ or both, and thus the finite-source effect with adaptive sampling is required. The largest deviation in this magnification map is $4.6 \times 10^{-4}$.
    Deviations below $10^{-7}$ are shown in white.
    \label{fig:mag_map}}
\end{figure*}
\section{Code Validation \& Performance }\label{sec:performance}

We use \vbbl\ (version = 3.6.2) as the benchmark to show the robustness and speed of \jaxcode\ in magnification calculations. The gradient results of \jaxcode\ are compared against the numerical gradients of \vbbl, which, as will be shown later, do not necessarily produce the correct results. A more robust validation of the gradient calculation is given in Section~\ref{sec:light_curve_analysis}, in which \jaxcode\ is applied to the analysis of a real light curve.

\begin{figure}[htb!]
    \includegraphics[width=\linewidth]{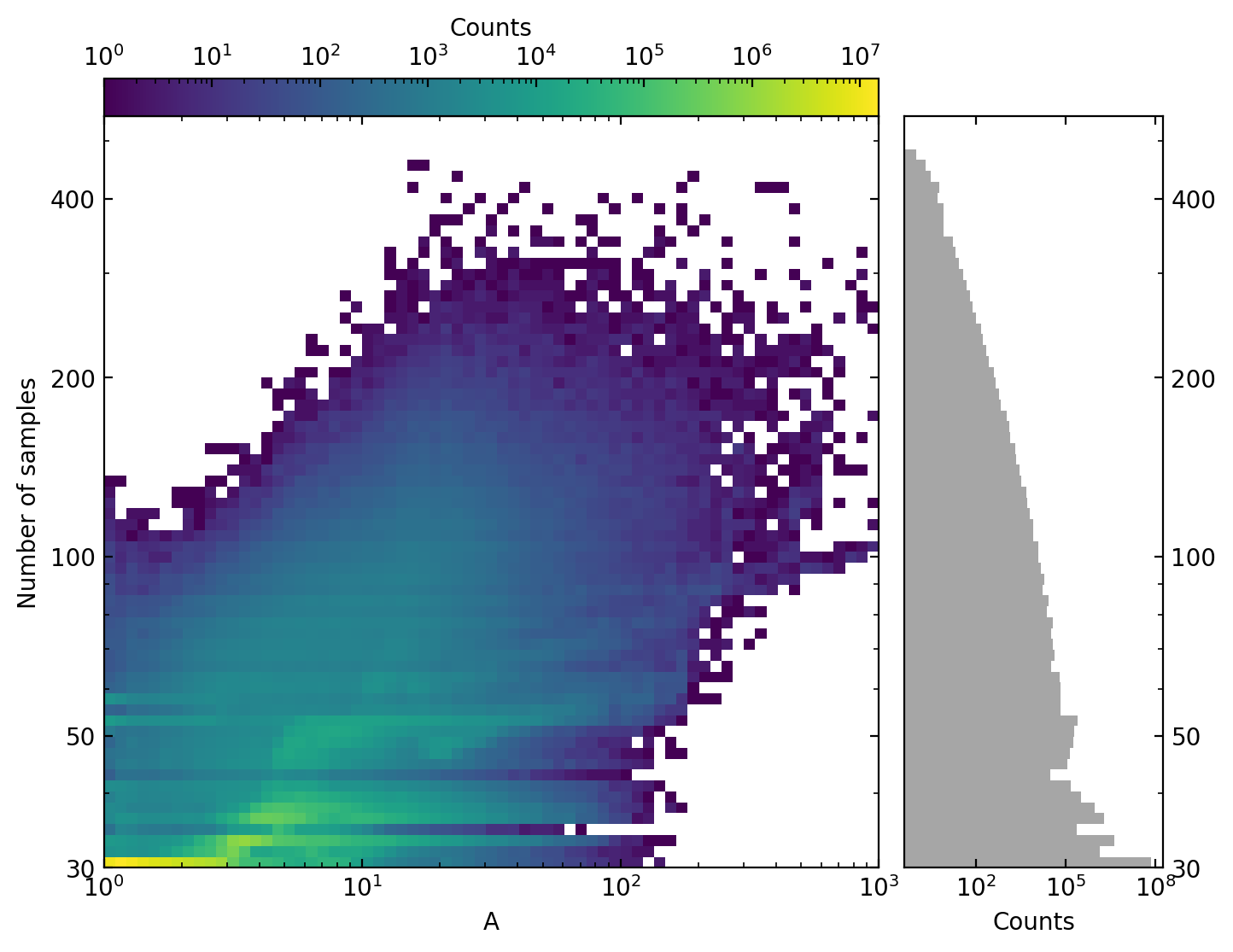}
    \caption{The left panel shows the total number of sample points used in the contour integration as the function of the microlensing magnification, color-coded by the number counts. The right panel shows the histogram distribution of the sampling points.
    }
    \label{fig:mag_vs_num}
\end{figure}

\begin{figure}[htb!]
    \includegraphics[width=\linewidth]{
    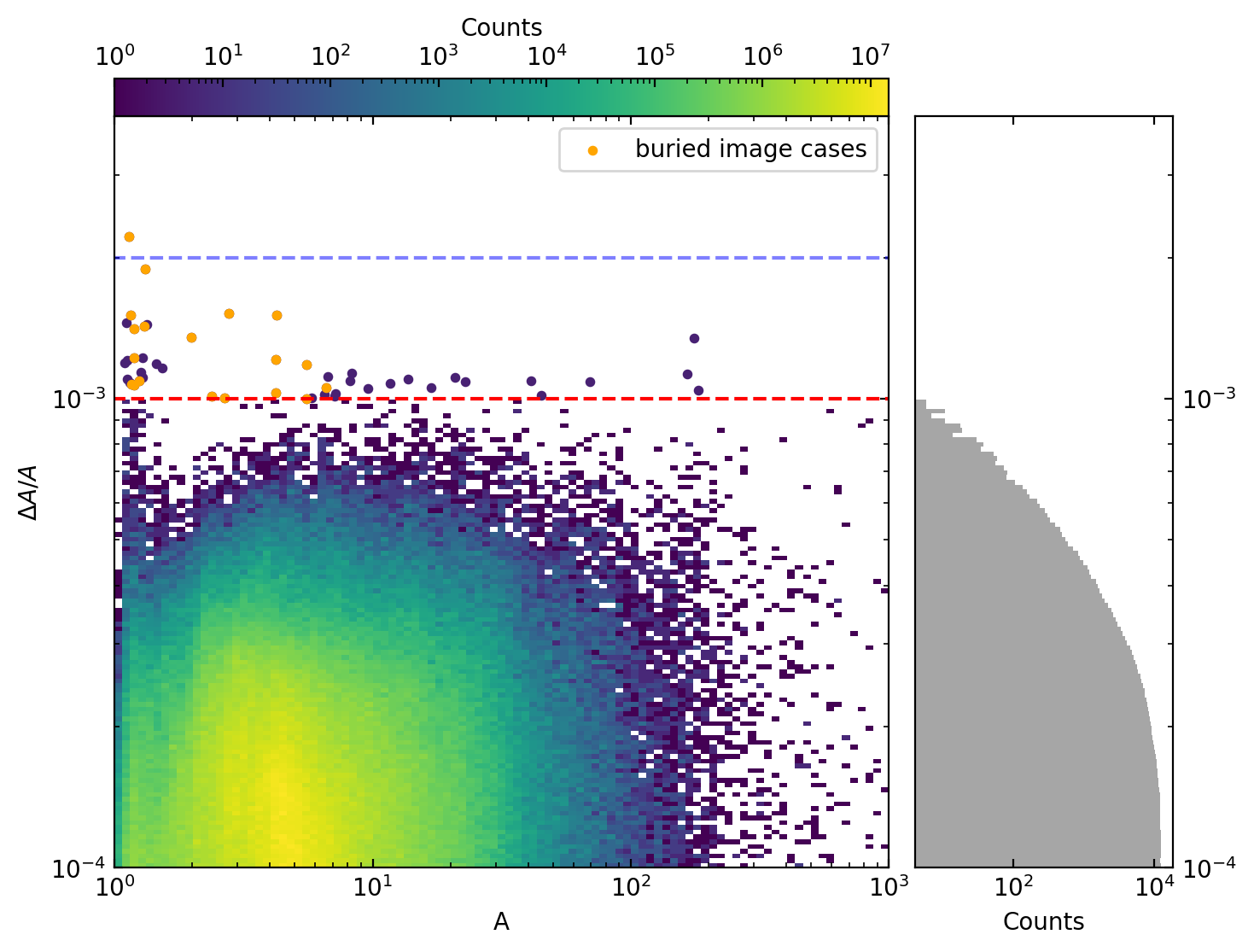}
    \caption{The left panel shows the fractional difference in microlensing magnification between \vbbl\ and \jaxcode, color-coded by the number counts. The red and blue dashed horizontal lines indicate the precision tolerance used in this test and twice of it, respectively. Of the $\sim8\times 10^7$ grid points shown here, 50 are above the red dashed line, the majority of which remain very close to the precision requirement. The maximum difference reaches $2.22 \times 10^{-3}$, due to the hidden cusp (see Figure~\ref{fig:hidded_cusp}). All cases (19 in total) that failed to reach the precision requirement due to the same reason are marked in orange. }
    \label{fig:mag_vs_rel_diff}
\end{figure}

\begin{figure}[htb!]
    \centering
    \includegraphics[width=0.95\linewidth]{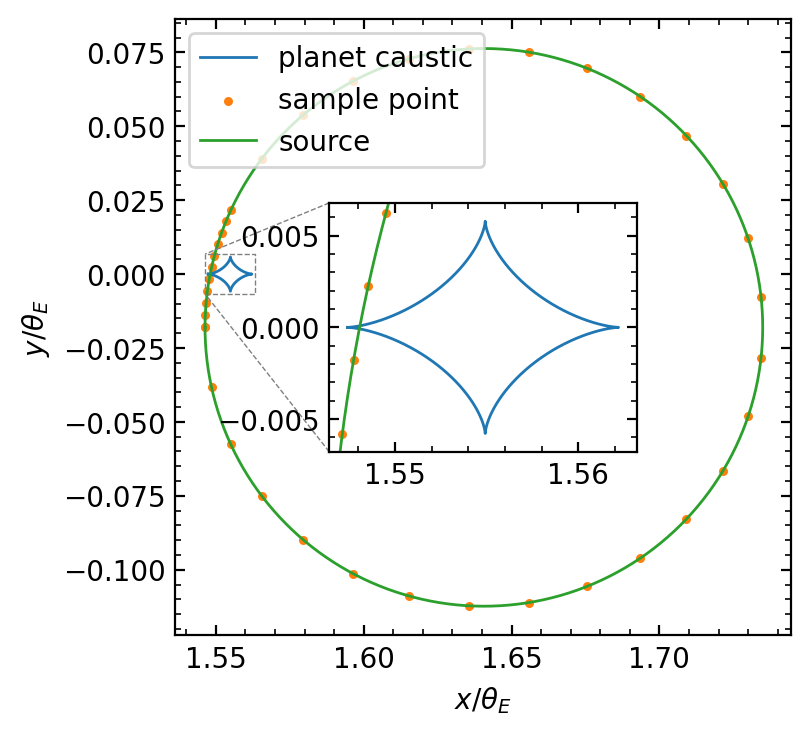}
    \caption{This figure shows the source star (green) and the caustic curve (blue) that correspond to the grid point in Figure~\ref{fig:mag_vs_rel_diff} with the largest deviation in microlensing magnification between \jaxcode\ and \vbbl. The key binary lens parameters are $\rho = 9.43 \times 10^{-2}, s = 2.04$, and $q = 1.81 \times 10^{-4}$. The orange dots are the sampled points according to the strategy of \jaxcode. The inset plot shows a zoom-in view around the intersection between the giant source and the tiny caustic. The cusp to the left is under sampled (i.e., a hidden cusp), which results in a relatively large deviation of $2.22\times 10^{-3}$ in the microlensing magnification. Such extreme cases are rarely encountered in realistic microlensing events.}
    \label{fig:hidded_cusp}
\end{figure}
\subsection{Microlensing Magnification} \label{sec:magnification_test}

To validate the magnification results of \jaxcode\ against \vbbl, we first generate 1000 combinations of binary lens parameters, with $\log q$, $\log s$, and $\log \rho$ randomly drawn from $(-6, 0)$, $(-0.5, 0.5)$, and $(-3, -1)$, respectively. For each parameter combination, we then use both \jaxcode\ and \vbbl\ to generate magnification maps with the dimensions of $1000\times1000$ in the range $-4<x<3 $ and $0<y<2$.
In these calculations, we have set the relative tolerance (i.e., $\Delta A/A$) to $10^{-3}$ for \jaxcode\ and $10^{-4}$ for \vbbl, which is used as the ground truth. We also increase the maximum length of the array related to adaptive sampling (Section~\ref{sec:bottleneck}) to 960 to avoid overflow.
One of the randomly generated magnification maps is shown in the left panel of Figure~\ref{fig:mag_map}, and the deviations relative to the similar map produced by \vbbl\ are shown in the right panel. The maximum deviation remains below the chosen tolerance, suggesting that the two codes produce very similar results at least in this particular example.

Out of $10^9$ grid points, about $8\times10^7$ fail the quadrupole test and thus require contour integration. This subset of grid points are used to investigate the computational efficiency and accuracy of our code. With our choice of the sampling strategy, the uniform sampling with 30 initial points is enough to meet the precision requirement for $\sim 86.7\%$ of the cases, whenever the contour integration is invoked. For the remaining $\sim 13.3\%$ of cases, additional samplings should be added via the adaptive sampling strategy in order to meet the precision requirement. The final number of sampling points are shown in Figure~\ref{fig:mag_vs_num} as a function of the microlensing magnification. Higher magnifications generally require more sampling points, although at any given magnification the exact number can vary substantially. Within the parameter space that is tested here, the maximum magnification is around 1000, and the maximum number of points needed is around 400, much below the maximum length size (960) used in these calculations.

The fractional difference in magnification is shown in Figure~\ref{fig:mag_vs_rel_diff} as a function of the magnification for all the grid points that require contour integration. All except for 50 of the $\sim 8\times 10^7$ grid points are below the precision tolerance of $10^{-3}$, indicating an overall good agreement between the two codes. Of the 50 grid points that exceed the precision tolerance, the majority (49/50) are still close to (i.e., below two times, indicated by the blue dashed line in Figure~\ref{fig:mag_vs_rel_diff}) the tolerance and considered acceptable.
\footnote{The error estimators contain a few parameters that are empirically determined. One can in principle choose to modify them to improve the precision, but we consider it unnecessary given that the difference remains small and acceptable.}
For the single grid point that exceeds twice the precision tolerance, the lensing geometry is shown in Figure~\ref{fig:hidded_cusp}, and the relatively large deviation comes from the under-sampled cusp region. As the deviation is still small, we do not further modify the code to account for such a rare case. Situations like this one, which have extremely large sources encountering tiny caustics, are also intrinsically rare in the analysis of realistic binary microlensing events.

\begin{figure*}[htb!]
    \includegraphics[width=\linewidth]{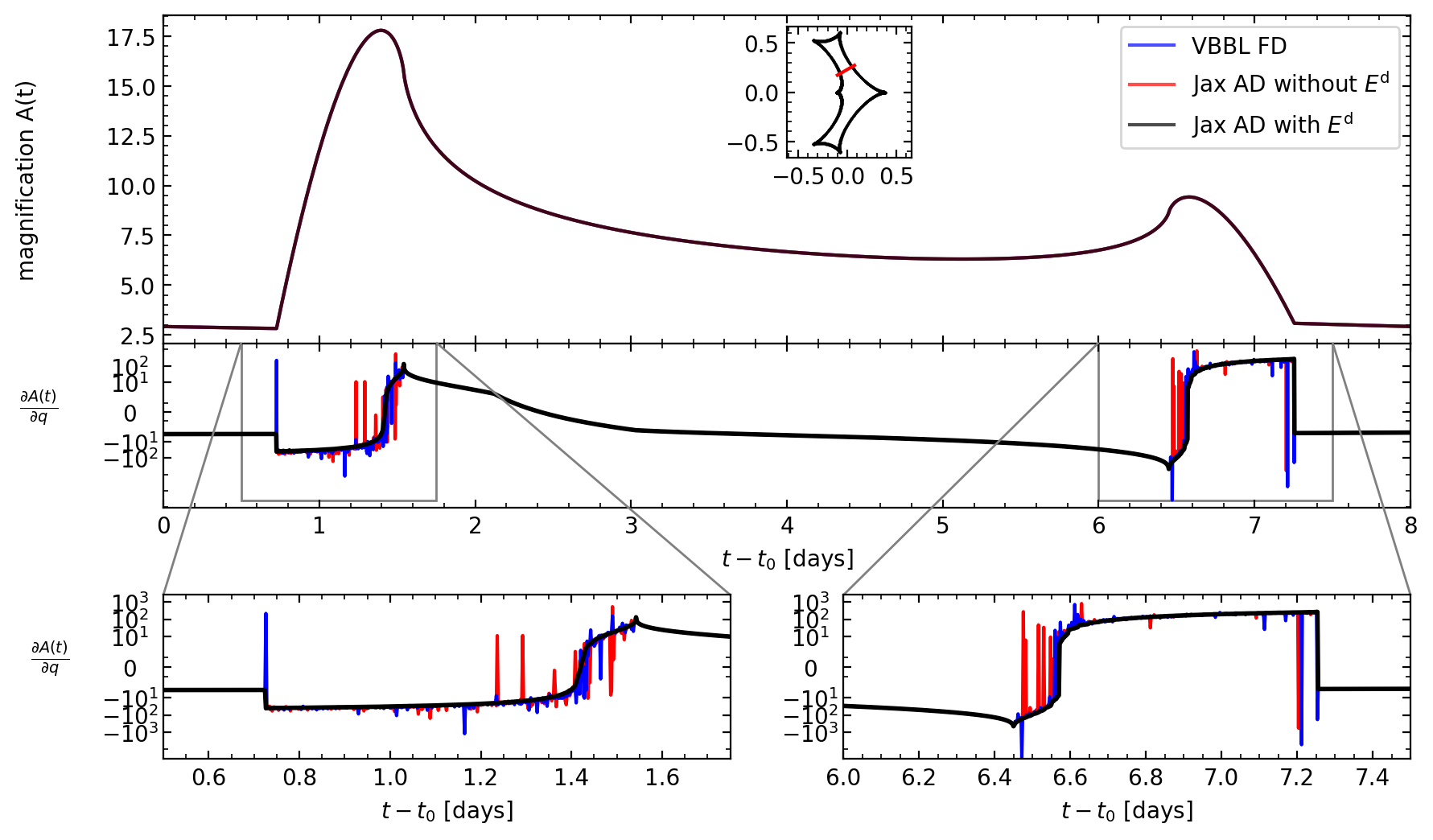}
    \caption{An example binary-lens light curve (top panel) and its derivative with respect to the planet-to-star mass ratio $q$ (middle panel). Three different methods have been used here, including using \jaxcode\ with (black curve) and without (red curve) the new error estimator term, $E^{\rm d}$, and using \vbbl\ through the finite difference method (blue curve; implemented in \texttt{scipy.approx\_fprime}). The bottom panels show the zoom-in views of the two caustic crossing regions. The inset in the top panel illustrates the binary-lens geometry (caustic and trajectory), with $q=0.2$, $s=0.9$, and $\rho=10^{-2}$. Only \jaxcode\ with $E^{\rm d}$ is able to yield smooth (and accurate) model gradient curve.}
    \label{fig:grad_comparison}
\end{figure*}

\begin{figure*}[htb!]
    \includegraphics[width=\linewidth]{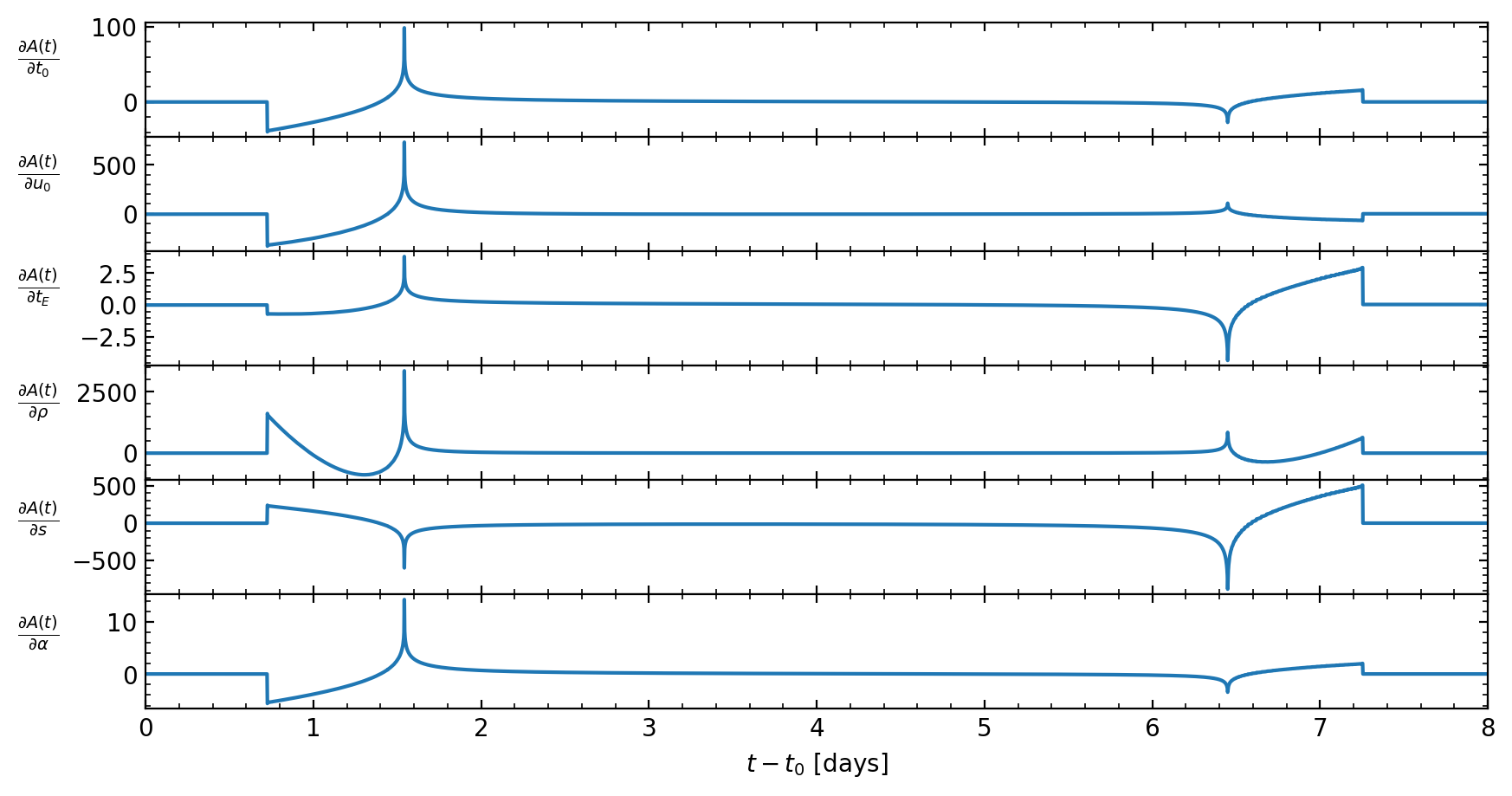}
    \caption{Derivatives of the same binary light curve as in Figure~\ref{fig:grad_comparison} respect to the other microlensing parameters, obtained by \jaxcode\ with the $E_4$ term.}
    \label{fig:grad_full}
\end{figure*}

Regarding the computational speed,
\jaxcode\ is 1.5--5 times slower than \vbbl\ in generating magnification maps with the same microlensing parameters and precision requirement. The exact slow-down factor depends on the fraction of grid points that fail the point source approximation and thus require contour integrals. For example, the computing time of the magnification map shown in Figure~\ref{fig:mag_map} is 22s for \vbbl, whereas it is 69s, thus a factor of 3 longer for \jaxcode\ for the same parameter setting. The above tests are done on a M2 MacBook pro with \texttt{Jax} version = 0.4.29. As the primary goal of \jaxcode\ is to provide accurate model gradients rather than to compete with \vbbl\ in the computation efficiency of microlensing magnifications, the speed difference of a factor of a few is sufficient for our purpose.
Nevertheless, we identify the primary causes of the speed difference as: (1) the less optimal sampling strategy (Section~\ref{sec:adaptive sampling in jax}) and (2) the less efficient polynomial solver (Section~\ref{sec:roots_finding}). Both have been chosen to increase the computational performance in \texttt{Jax}. Additionally, the use of arrays with fixed lengths (Section~\ref{sec:bottleneck}) and
the less efficient data movement in arrays compared to linked lists also contribute to the speed difference. Some of these issues are the subjects of future optimization of our code.%

\subsection{Gradient} \label{sec:gradient}
We use \texttt{jax.jacfwd} to obtain the model gradients. An example is given in Figure~\ref{fig:grad_comparison}, which shows the microlensing light curve and its partial derivative with respect to the planet-to-star mass ratio $q$, $\partial A/\partial q$. The automatic differentiation of \jaxcode\ with the addition of the new error estimator $E^{\rm d}$ (Equation~\ref{eqn:new-error}) gives a smooth curve for the chosen gradient. For comparisons, the result of \jaxcode\ without $E^{\rm d}$ and the gradient calculated by \vbbl\ based on finite difference both produce erroneous gradient curves, evidenced by the numerous spikes during the caustic crossing regions. In calculating the gradient with \vbbl\ via the finite difference method, we have used \texttt{scipy.approx\_fprime} with the default step size ($\delta \rho\approx 1.5 \times 10^{-8}$) and set the precision tolerance to $\Delta A/A=10^{-3}$. The accuracy in the gradient can in principle be improved with a much lower tolerance, but that would substantially reduce the speed of \vbbl\ with no guarantee on the accuracy of the gradient. %
For completeness, we show the derivatives of the same light curve model with respect to the other model parameters in Figure~\ref{fig:grad_full}.  

Our code provides accurate model gradients in a computationally efficient way.
The computational time of the seven gradients and the magnification value altogether is only 2--3 times longer than the computational time of the magnification value alone.\footnote{For the specific example shown in Figures~\ref{fig:grad_comparison} and \ref{fig:grad_full}, they are 4.09s and 1.47s, respectively.}
For comparisons, the finite difference method will need eight function evaluations to obtain the gradients of seven parameters, in addition to its incapability to yield accurate gradients.

\section{Application to a real event} \label{sec:light_curve_analysis}
Access to accurate model gradients allows us to use gradient-based methods in the search for best-fit model parameters and the estimation of the parameter posteriors. Here we demonstrate the application of \jaxcode\ to the second task while leave the former to some future work.

We select the microlensing event KMT-2019-BLG-0371 as the example to demonstrate the utility of gradients in microlensing analysis and the robustness of our code. KMT-2019-BLG-0371 is a relatively short-timescale ($t_{\rm E}=6.65$) binary-lens event with $q=7.24\times 10^{-2}$ and prominent caustic-crossing features \citep{Kim2021KMT2019BLG0371}. This event has degenerate solutions, but here we only focus on the small $s$ solution. We assume that the best-fit parameters have been identified and focus on the estimation of the parameter posteriors.

Besides the seven parameters that are needed to describe the microlensing magnification $A(t)$ at a given time $t$, two flux parameters for each data set are usually needed to model the actual microlensing light curve
\begin{equation} \label{eqn:fsfb}
    F(t) = F_{\rm S} \cdot A(t) + F_{\rm B} .
\end{equation}
Here $F_{\rm S}$ is the flux of the source star in the absence of lensing, and $F_{\rm B}$ is the blending flux. For simplicity, below we use the aligned light curve and thus only one set of $F_{\rm S}$ and $F_{\rm B}$ is needed. As Equation~(\ref{eqn:fsfb}) is linear, the derivatives of the model flux with respect to all model parameters (i.e., the seven magnification parameters and the two flux parameters) can be easily derived, given the gradients of $A(t)$ from \jaxcode.

\begin{figure*}[htb!]
    \plotone{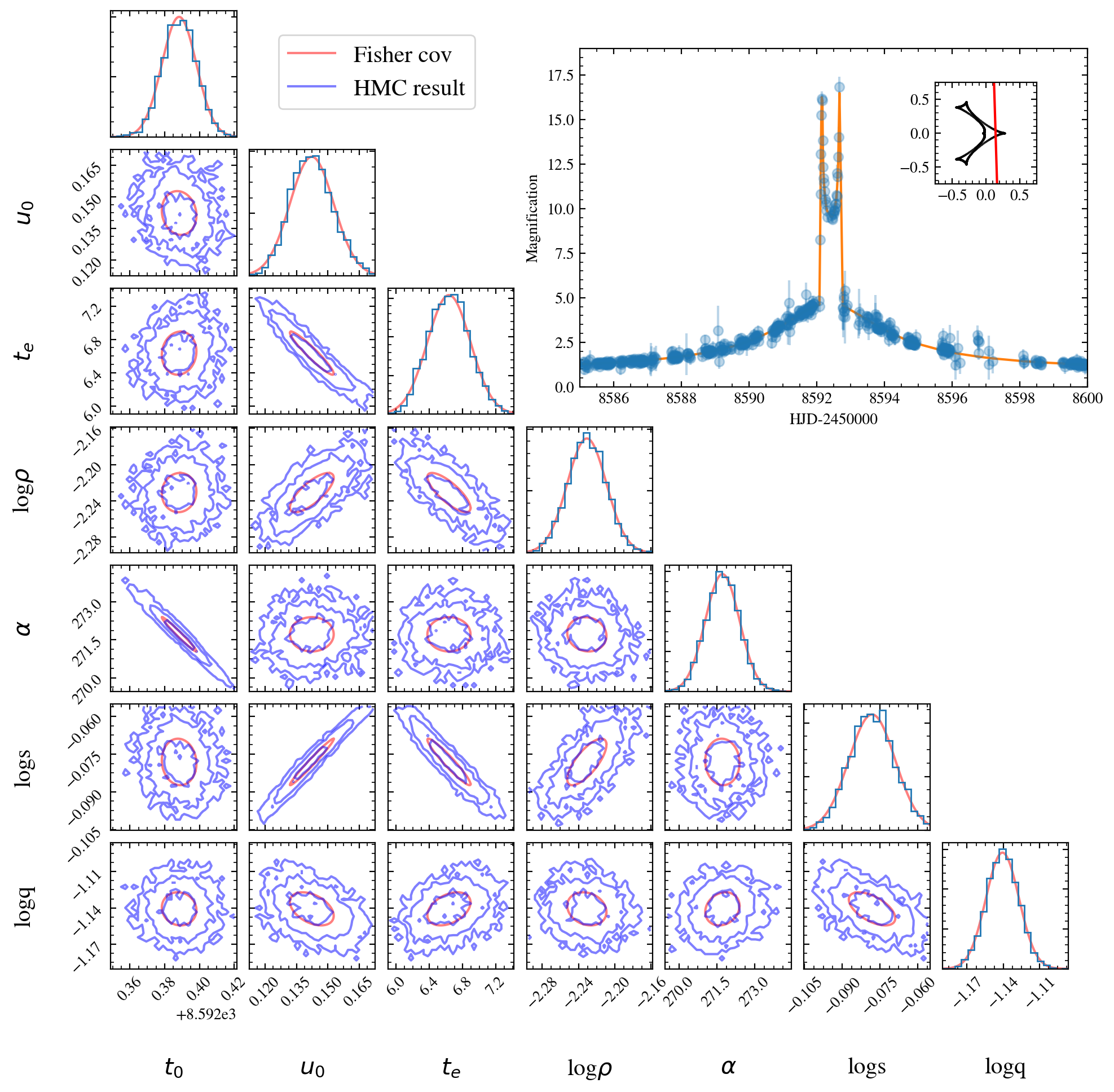}
    \caption{The corner plot illustrating the covariance matrices of KMT-2019-BLG-0371 from the Fisher information matrix approach (red curve) and the HMC sampling (blue curve). Only the 1-$\sigma$ contours are shown in the former case, whereas the 1-3$\sigma$ contours are shown in the HMC case. The light curve and the binary-lens geometry for the close solution are shown in the upper right figure.}
    \label{fig:fisher&HMC}
\end{figure*}

\subsection{Fisher Information Matrix} \label{sec:fisher}

For binary lens events, the model posterior can often be described as a multivariate Gaussian.
Once the best-fit parameters are known, the covariance matrix of the Gaussian is given by the inverse of the Fisher information matrix,
which is defined as 
\begin{equation}
\mathcal{F}
_{ij} =\sum_k \frac{1}{\sigma^2_{k}}{\frac{\partial F(t_k)}{\partial \Theta_i }} {\frac{\partial F(t_k)}{\partial \Theta_j }} .
\end{equation}
Here $\Theta_i$ is the $i$-th model parameter, and $F(t_k)$ and $\sigma_k$ are the model light curve and the observed flux uncertainty at epoch $t_k$, respectively. In Figure~\ref{fig:fisher&HMC}, we compare the covariance matrix derived from the Fisher information matrix and that from the sampling method (see Section~\ref{sec:hmc}), and the agreement is excellent. It takes only one function evaluation to obtain the covariance matrix via the Fisher information approach. 

The model posterior may deviate, sometimes substantially, from a multivariate Gaussian. This usually happens when the correlations between certain model parameters are strong and/or the constraints on certain parameters are weak. In the first case, a multivariate Gaussian can still be achieved and the Fisher information approach remains applicable as long as the model can be reparameterized in a proper way \citep[e.g.,][]{Zhu:2015}. In the latter situation, the Fisher information approach is still useful in estimating the covariances of those well-constrained parameters. It also serves as a useful starting point for the sampling approach to map the posterior in a more efficient way (see Section~\ref{sec:hmc}).

\subsection{Hamiltonian Monte Carlo} \label{sec:hmc}

Monte Carlo sampling methods are often used to map the model posterior. The access to model gradient of \jaxcode\ allows us to employ more efficient Monte Carlo methods other than the commonly used standard Markov Chain Monte Carlo \citep[MCMC, e.g.,][]{foreman-mackey2013}.

Here we use Hamiltonian Monte Carlo \citep[HMC,][]{duaneHybridMonteCarlo1987, neal2011} to showcase the potential of \jaxcode\ in efficiently sampling the posterior distribution. Specifically, we choose the No U-Turn Sampler \citep[NUTS,][]{hoffman2011} in \texttt{Numpyro} \citep{bingham2019pyro, phan2019composable}. 
To achieve better performances, we use the covariance matrix obtained from the Fisher information matrix approach (Section~\ref{sec:fisher}) to reparameterize the binary microlensing parameters, which largely removes the correlations between parameters and thus improves the efficiency in the warm-up stage in NUTS. We run 4 chains, each with 500 warm-up steps and 1000 sampling steps. The sampler works successfully with an average acceptance rate of 0.88 
. The resulting posterior distribution is also shown in Figure~\ref{fig:fisher&HMC}. 

To fairly compare the performance of different samplers, we use Effective Sample Size (ESS) as the criterion \citep{geyer2011}. For the same dataset and initial condition, we obtain an ESS of 504 when using \texttt{emcee} with 40 chains and 2000 samples in each chain, yielding 
158 function calls per ESS
. For comparisons, NUTS combined with \jaxcode\ obtains an ESS of 1726 based on 56712 function calls, thus about 33 function calls per ESS.
Therefore, NUTS is more efficient than \texttt{emcee} by a factor of about five in this specific example.
\footnote{In terms of the speed, the difference in this case is also about five, with NUTS being faster. This is because the parallelization in \texttt{Jax} is more efficient than that in \texttt{emcee}, even though \jaxcode\ is slower than \vbbl\ by a factor of a few.}
The difference in sampling efficiency is expected to be even larger once the posterior distribution becomes more complicated and/or the higher-order effects such as microlensing parallax and lens orbital motion are included.
We leave a more detailed investigation of this to a future work.

\section{Discussion}\label{sec:discussion}

We present \jaxcode, \footnote{\href{https://github.com/CoastEgo/microlux}{https://github.com/CoastEgo/microlux}}
which implements a modified version of the adaptive contour integration method of \citet{bozza2010} in the \texttt{Jax} library. This differentiable code can achieve efficient and accurate calculations of binary microlensing light curve and its gradient. Compared to the existing codes, a few notable differences are introduced in order to optimize the performance of \jaxcode\ in \texttt{Jax}. These include the different algorithms to solve the lens equation and connecting the images, a different sampling strategy, and the use of dynamic data structures.

\jaxcode\ is also specially designed to obtain accurate model gradients.
In particular, we have introduced a new error estimator into the adaptive sampling algorithm in order to ensure the convergence of the gradient.
We have tested the efficiency and accuracy of \jaxcode\ against the existing code in the relevant parameter space of binary microlensing modeling. In its current implementation, \jaxcode\ is slower by a factor of $<5$ in computing the microlensing magnification, but can provide faster and more accurate evaluation of the model gradients.

The access to accurate model gradients can potentially speed up the modeling process of binary microlensing events. In Section~\ref{sec:light_curve_analysis}, we have applied \jaxcode\ to a real binary microlensing event to showcase its power in estimating the model posterior. Once the best-fit solution is known,
\footnote{Although not demonstrated here, the search for the best-fit solution also benefits from access to the model gradients.}
the posterior distribution can be obtained via the inversion of the Fisher information matrix, which only requires one function evaluation, or the advanced sampling method that utilizes the model gradients (e.g., HMC), which can achieve faster convergence than the standard MCMC approach. In addition, the differentiable nature of \jaxcode\ may also improve the performance of machine learning models in predicting the parameters as well as their posteriors of binary microlensing events \citep{Zhang:2021, Zhao:2022}.

Finally, it is worth pointing out that there is still room for further optimization in \jaxcode. As noted in Section~\ref{sec:magnification_test}, the sampling strategy and the polynomial solver are most responsible for the speed difference between \jaxcode\ and \vbbl. In the current implementation, these two steps have been designed to balance the robustness and efficiency across the wide range of parameter space relevant for binary microlensing, they can certainly be further optimized if specific parameter ranges are concerned. For example, the semi-analytical root solver proposed by \citet{zhang2023} may perform better for typical planetary microlensing events. A more efficient use of the dynamical array and the implementation into GPU \citep{wang2025} may potentially speed up the code as well. We leave the investigation of these possibilities to some future work.

\begin{acknowledgments}
The authors thank Valerio Bozza, Renkun Kuang, Kento Masuda, Jiyuan Zhang, Keming Zhang and Haimeng Zhao for the discussions and suggestions about the code development. 
This work is supported by the National Natural Science Foundation of China (grant Nos.\ 12133005 and 12173021).
The authors acknowledge the Tsinghua Astrophysics High-Performance Computing platform at Tsinghua University for providing computational and data storage resources that have contributed to the research results reported within this paper.
\end{acknowledgments}

\vspace{5mm}

\software{Numpy \citep{numpy2011,numpy2020}, Scipy \citep{scipy2020}, Matplotlib \citep{matplotlib2007,matplotlib_2016}, Jax \citep{jax2018github}, VBBinaryLensing \citep{bozza2010,bozza2018}, emcee \citep{foreman-mackey2013}, corner \citep{corner}, Numpyro 
 \citep{bingham2019pyro, phan2019composable}, BlackJax \citep{cabezas2024blackjax}, Jupyter \citep{jupyter}          }

\bibliography{sample631}{}
\bibliographystyle{aasjournal}

\end{document}